\def \figurecaption            {\noindent\hangindent0.5in\hangafter=1}
\def \bullit         {\ifmmode \bullet \else $\bullet$ \fi}
\def \deg            {\ifmmode ^\circ \else $^\circ $ \fi}   

\def \mp             {\ifmmode \pm \else $\pm$ \fi}

\def \ref            {\noindent\hangindent0.5in\hangafter=1}

\def \um             {\ifmmode \mu m\else $\mu$m\fi}
\def \x              {\ifmmode \times \else $\times$ \fi}
\def \HOH            {\ifmmode H_2O\else H$_2$O\fi}
\def \HH             {\ifmmode H_2\else H$_2$\fi}
\def \OO             {\ifmmode O_2\else O$_2$\fi}
\def \oz             {\ifmmode O_3\else O$_3$\fi}
\def \Oplus          {\ifmmode O^+\else O$^+$\fi}
\def \Hplus          {\ifmmode H^+\else H$^+$\fi}
\def \Splus          {\ifmmode S^+\else S$^+$\fi}
\def \SOt            {\ifmmode SO_2\else SO$_2$\fi}
\def \ch             {\ifmmode CH_4\else CH$_4$\fi}
\def \nh             {\ifmmode NH_3\else NH$_3$\fi}
\def \CQ             {{1997~CQ$_{29}$}}
\def \WW             {{1998~WW$_{31}$}}

\def \CF             {{2000~CF$_{105}$}}
\def \TC             {{1999~TC$_{36}$}}
\def \SM             {{1998~SM$_{165}$}}
\def \QT             {{2001~QT$_{297}$}}
\def \QW             {{2001~QW$_{322}$}}
\documentstyle[aasms4]{article}
\begin{document}

\title  {Detection of Two Binary Trans-Neptunian Objects, \CQ\ and \CF ,
with the Hubble Space Telescope}

\author{ Keith S.~Noll, Denise C.~Stephens}
\affil{Space Telescope Science Institute}

\author{ Will M.~Grundy, Robert L.~Millis, John Spencer, Marc W.~Buie}
\affil{Lowell Observatory}

\author{Stephen C.~Tegler}
\affil{Northern Arizona University}

\author{William Romanishin }
\affil{University of Oklahoma}

\author{Dale P.~Cruikshank}
\affil{NASA Ames Research Center}

\begin{abstract}  

Images of the trans-Neptunian objects \CQ\ and \CF\ obtained with the
Hubble Space Telescope's WFPC2 camera show them to be binary.  The two
components of \CQ\ were separated in our images by 0.20$\pm$0.03 arsec
in November 2001 and by 0.33$\pm$0.01 arcsec in June/July 2002.  The 
corresponding minimum physical distances are 6100 km and 10,200 km.  The
companion to \CF\ was 0.78$\pm$0.03 arcsec from the primary, at least
23,400 km.  Six other objects in the trans-Neptunian region, including
Pluto and its moon Charon, are known to be binaries; \CQ\ and \CF\ are
the seventh and eighth known pair.  Binarity appears to be a
not-uncommon characteristic in this region of the solar system, with
detectable companions present in 4$\pm$2\% of the objects we have
examined.

\end{abstract}

\keywords{Kuiper Belt, Oort cloud}

\section{Introduction}

The first trans-Neptunian binary (TNB) was identified in 1978 with the
discovery of Charon, Pluto's moon (Christy and Harrington 1978).  The
first binary Kuiper-belt object, \WW , was announced in 2001 (Veillet
2001), though it has since been identified in images taken as early as
November 1998 (Veillet et al. 2002).  More discoveries have followed in
rapid succession (Table 1) so that, as of February 2002, a total of
six trans-Neptunian binaries were known.

The discovery of TNBs opens up a significant new tool for physical study
of trans-Neptunian objects (TNO).  Through analysis of the orbit, it
offers the only direct means of determining the mass of these distant
objects for the forseeable future.  When combined with optical and
thermal wavelength photometry, density can also be determined.  Both
mass and density have {\it a priori} uncertainties of at least an order of
magnitude for TNOs, so the value of direct observations cannot be
underestimated. 

Binaries offer further opportunities for physical study through mutual
occultations.  The series of Pluto-Charon mutual events that occurred
soon after the discovery of Charon led to surface albedo maps and
separate spectra, revealing the very distinct surface compositions of
Pluto and its satellite (Buie et al. 1987, Fink \& DiSanti 1988, Buie
et al.~1992, Young et al.~1999).  Veillet et al.~(2002) suggest that
\WW\ will begin a series of mutual events in approximately 50 years. 
TNOs have periods of 250 years or more so it will be necessary to
discover and determine the orbits of approximately 10 TNBs before it
will be likely to find one having mutual occultations within the next
decade.  

Two TNOs, \CQ\ and \CF , are the focus of this work.  \CQ\ was
discovered at Mauna Kea Observatory by Chen et al. (1997) and \CF\ was
discovered at Kitt Peak National Observatory by the Deep Ecliptic
Survey team (Millis, et al. 2002).  Both objects are classical TNOs in
orbits of small eccentricity and low inclination.  In this paper we
describe the discovery of binary companions to \CQ\ and \CF , making
them the seventh and eighth known TNBs (Noll and Stephens 2002, Noll et
al. 2002).

\section{Observations}

\CQ\ and \CF\ were observed as part of a large photometric survey of
TNOs that we are carrying out using the Hubble Space Telescope.    The
data were dark-subtracted, flat-fielded, and flux calibrated using
standard STScI on-the-fly pipeline calibration steps including
up-to-date dark files.  Cosmic ray hits were removed by combining pairs
of filtered observations with the standard CRREJ routine found in the
STSDAS IRAF software package (Baggett et al.~2002).  

We used the WFALL aperture, centered at pixel 133, 149 on the WF3 chip
(Biretta et al.~2001), for all observations in our photometric survey. 
This is the corner of the WF3 chip nearest the x and y readout
registers, a position that reduces the magnitude of the charge transfer
efficiency correction required (Whitmore et al. 1999, Dolphin 2000). 
The WFALL aperture was also chosen to increase the likelihood of
finding objects with uncertain ephemeris positions.  The telescope was
tracked during all observations correcting both for the proper motion
of the TNO and the parallax induced by Hubble's orbital motion.  

In followup observations of \CQ\ we placed the target on the PC chip to
take advantage of the better sampling on that detector.  As with the
other observations, the telescope was tracked at the apparent rate of
the TNO.

\subsection{\CQ }

\CQ\ was first imaged by us from 6:44 to 7:22 UT on 17 November 2001
when it was at a distance of 41.85 AU from the Earth.  We obtained
images with the  WFPC2 camera on the Hubble Space Telescope (HST) in
three broad band filters approximating V (F555W), R (F675W), and I
(F814W) bands. There were two 160 second exposures in each filter.  The
sequence started with a single F555W exposure, followed by two each in
F675W and F814W, and ended with a second single F555W.  

We identified \CQ\ in our images both by its lack of motion relative to
tracking and by its position at 66,92 on the WF3 chip, 8.76 arcsec from
the predicted position.  Stars and galaxies in the field moved by a
total of 0.46 arcsec between the first and last exposure.

As shown in figure 1, the image of \CQ\ is clearly elongated.  The
orientation of the spacecraft was such that north is approximately 30
degrees from the vertical in our images as indicated by the long arrow.
 The axis of the elongation is rotated by 16$\pm$4 degrees east of
north relative to the brighter of the two components, component $A$. 
The elongation appears in all six individual images at a similar angle
and separation.  We note that the direction of the elongation is almost
perpendicular to the apparent motion of stars and galaxies which moved
diagonally from east to west in the images as shown.  We also examined
engineering data that tracks jitter during the observations.  No
unusual jitter was seen, ruling this out as a possible source of error.
Based on this visual evidence, we concluded that \CQ\ is a probable
binary and proceeded to obtain confirming images.

Three additional sets of observations of \CQ\ were made from 7:30 to
8:26 UT on 18 June 2002, 3:09 to 4:05 UT on 30 June 2002, and 3:39-4:35
UT on 12 July 2002.  On each date we obtained three 800 second exposures
in the F814W filter with the \CQ\ on the PC chip.  In all three cases
the binary was easily detected and clearly resolved, as shown in figure 1.
The higher resolution of the PC, the better S/N resulting from longer
integrations, and obvious presence of the two components on four
separate dates leaves no doubt about the binarity of \CQ .

\subsection {\CF }

\CF\ was observed in two filters, the F555W and F814W, on 12 January
2002 from 1:12 to 1:49 UT.  As with \CQ , the observing sequence began
and ended with an F555W exposure.  A pair of F814W exposures occurred
between the F555Ws.  Because \CF\ is fainter, we exposed this object in
only two filters with two 260 second exposures per filter.

The brighter component of \CF , component $A$, was found 0.58 arcsec
from the WFALL aperture at pixel 133,155.  A second, fainter object,
component $B$, can
be seen to the east in figure 2 at pixel 138, 161.  \CF\ $A$ and $B$
were stationary during the observations.  Stars and galaxies
were trailed, moving a total of 1.1 arcsec from the first to last
exposure.  As shown in figure 2, the two components of \CF\ are cleanly
separated by an angular distance of 0.78$\pm$0.03 arcsec at a position
angle of 106.6$\pm$2.5 degrees measured from $A$.   

\CF\ was subsequently observed from the Keck I telescope (Fig.~3,
Romanishin et al.~2002).  On the night of 11 April 2002 (UT) \CF\ was
observed in four ten-minute integrations simultaneously in B and R
bands using the LRIS instrument in imaging mode.  The images were
individually bias-subtracted and flat-fielded and then combined into a
single image.  For the combined image shown in figure 3, the individual
exposures were shifted to align the images of \CF\ which was moving at
-0.235 arcsec/hr in RA and 0.00 arcsec/hr in dec.  As can be seen in
figure 3, \CF\ appears elongated in the east-west direction with the
fainter component to the east.  Romanishin et al. report a separation
of 0.8$\pm$0.2 arcsec at a position angle of 103$\pm$5 degrees.  This
matches the orientation seen in the WFPC2 image and confirms the
presence of the companion.

\section{Analysis}

\subsection{\CQ }

In order to determine the separation and to obtain separate photometric
magnitudes for the two components of \CQ\ in the November data we fit
the observed image by iteratively combining scaled synthetic
point-spread functions (PSF) generated by the TinyTim software tool
(Krist et al.~1997).  An example is shown in figure 4.  Both the
separation and the relative magnitude of the components were allowed to
vary.  After finding a preliminary fit, we repeated the procedure over
a finer grid.  A best fit was determined by finding the parameters that
yielded a minimum in the least squares residuals between the model and
observed data.  Errors were estimated from the residuals.

From the iterative fit we determined a mean separation of 0.20$\pm$0.03
arcsec on 17 November 2001.  We note that this is 0.04 arcsec larger
than reported by Noll \& Stephens (2002) because the separation of the
PSF centers is slightly larger than the separation of the peaks in the
brightness distribution.  This separation is near the resolution limit
of the wide field portion of WFPC2 where the pixel scale is 0.1
arcsec/pixel, though it is sufficiently large that the two components
can be reliably resolved.   We applied the same technique to determine
separations in the June/July 2002 data, even though the objects are
clearly resolved and separate.  Positional information is summarized in
Table 2.

Magnitudes are converted to the standard Johnson-Cousins system using
the SYNPHOT software package.  Details of our photometric pipeline are
described in more detail in a forthcoming paper.  We measure a Johnson
V magnitude of V = 23.78$\pm$0.13 and Cousins R and I magnitudes of R =
23.27$\pm$0.19,  I = 22.54$\pm$0.09 for the component $A$.  Component
$B$ was found to have V = 24.02$\pm$0.22, R = 23.54$\pm$0.23, and I =
23.06$\pm$0.18 (Table 3).  The photometry we obtained in June/July 2002
is generally consistent with the I band magnitudes determined in
November 2001 as detailed in Table 3.  

The one exception is the unusually faint magnitude recorded for the
component $A$ on 18 June 2002.  The faintness initially caused us to
misidentify the components in this image, but when combined with the
other images and possible mutual orbits, it is clear that on 18 June
\CQ $A$ is almost a full magnitude fainter than on the other three
dates when we measured it.  Note, in particular, that the observations
in June and July 2002 were each separated by approximately 12 days. 
This is far shorter than any plausible period for the binary as
discussed below.  And there is essentially no change between the 30 June
and 12 July magnitudes for either component.  Finally, the measured
magnitude of component $B$ is almost identical in all four observations.
 It is possible that the unusually low measured magnitude for component
$A$ is due to some unrecognized observational error.  It is also
possible that \CQ $A$ has an intrinsic lightcurve with a large
amplitude.  Further observations to set limits on lightcurves would be
valuable.

The discovery of binaries is valuable because it offers the possibility
of determining the mass and density of the objects, basic physical
parameters that would otherwise require spacecraft to measure.  We have
not yet determined the orbits for either of the binaries that we have
detected.  It is possible, however, to make estimates of diameter and
mass by making assumptions about the albedo and density of the objects
and using measured magnitudes.  If we assume an albedo of q=0.04 as is
customarily done for TNOs, we find that the two components of \CQ\ have
diameters of d(A)=220 km  and d(B)=200 km following the formalism of
Romanishin and Tegler (1999).  If we further assume a bulk density of 1
g cm$^{-3}$, we calculate a combined mass of 9.76$\times 10^{21}$ g, or
1/1500th the mass of Pluto and Charon.  It is also possible to make
educated guesses about possible orbital periods as we discuss separately
below. 

%
%
%
%
%
%
%

\subsection{\CF }

On January 12, 2002, \CF\ was 41.30 AU from the Earth.  At this
distance, the angular separation we observed, 0.78 arcsec, corresponds
to a minimum separation of 23,360 km.   Because of the wide separation,
we were able to measure individual magnitudes for each component of the
binary with standard aperture photometry using the IRAF PHOT routine. 
We used an aperture radius of 3 pixels and a background annulus with
inner diameter of 20 pixels and a thickness of 20 pixels.  An aperture
correction factor was calculated using synthetic PSFs.  We determined V
= 24.25$\pm$ 0.11 and I = 22.99$\pm$ 0.08 for the brighter component
$A$ and V = 25.12$\pm$ 0.24 and I = 23.55$\pm$ 0.13 for the component
$B$ (Table 3).  The uncertainties are relatively large, especially for
the fainter component $B$, because of the faintness of these objects. 
\CF $A$ is one of the faintest TNOs we have observed and the
\CF $B$ is more than a magnitude fainter than the nominal magnitude
limit we used for target selection.  Significantly fainter companions
would not be detected by our relatively short integrations.

Romanishin et al.~(2002) report that the two components of \CF\ differ
by 0.6$\pm$0.2 mags in the R band, consistent with HST relative
photometry.  Absolute photometry with the Keck data has not been
reported.

Using the same procedure as we did for \CQ , we estimate diameters for
the two components to be d($A$)=170 km and d($B$)=120 km for an albedo of
0.04.  Assuming a density of $\rho$ = 1 g cm$^{-3}$ we find a combined
mass of 3.48$\times 10^{21}$g, 4200 times less than Pluto and Charon.

\subsection {Orbits}

It is possible to estimate the period of TNBs using Kepler's third law
by making an assumption about the semimajor axis and eccentricity of
the orbit.  This exercise is necessary in order to plan followup
observations and is also useful as a yardstick for comparison with the
actual orbit once it is known.  We note that of the two TNBs with
measured orbits, one, Pluto/Charon is nearly circular, while the other,
\WW\ has an eccentricity of e=0.8 (Veillet et al. 2002).

For our initial guess we assumed that e=0 and that the observed
separation in November 2001, a=6070 km, is the semimajor axis.  For
this set of assumptions we derive an orbital period of 42.6 days.  We
used this value to plan our followup observations in June and July
2002.  The separation we observed in June/July 2002 was larger, an
average of 10,200 km.  Using the same assumptions, this would yield a
period of 92 days.  

The four epochs of observation allow us to determine some basic facts
about the orbit of \CQ .  As can be seen in Table 2, on the dates in
June and July 2002, each separated by approximately 12 days, the
position angle measured from component $A$ to $B$ increases steadily. 
The average rate over the nearly 24 days from 18 June to 12 July 2002
is 0.281 degrees/day.  However, the average rate from 17 November to 18
June, a period of 212.85 days, must be significantly higher, averaging
1.494 degrees/day.  We have insufficient data to resolve these
discrepant rates.  It is apparent, however, that the orbital plane is
inclined relative to our line of sight and/or the orbit is
significantly eccentric.  Additional observations will be required to
determine the orbit of \CQ .

If we perform the same exercise for \CF\ we find an orbital period for
a circular orbit would be 539 days, comparable to the 547 day period of
\WW\ (Veillet et al.~2002) and significantly longer than the apparent
period of \CQ .  The similarity of position angles observed in January and
April 2002 is consistent with a long period, though the uncertainties
are sufficiently large that nothing more quantitative can be said.

\subsection{Colors }

The V-I color index can be computed for each binary component from the
measured V and I photometry.  Color can, in principle, be used to infer
the surface composition and resurfacing history of TNOs (c.f. Jewitt \&
Luu 2001).  In the 75 objects surveyed by us, V-I ranges from 0.8 to
1.6, which is to be compared to V-I(Sun)=0.71. That is, TNO colors
range from neutral to extremely red, as has been noted before (Luu \&
Jewitt 1996).  

The colors of \CQ\ and \CF\ are listed in Table 3.  The brighter $A$
components of both binaries have colors, V-I $\sim$ 1.25, that are
average for classical TNOs observed in our survey.  Their companions
differ by about 1-sigma, more neutral in the case of \CQ $B$ and redder
in the case of \CF $B$.  Unfortunately, the uncertainties in the colors
are sufficiently large that we cannot reach conclusions beyond stating
that \CF $B$ is unlikely to have neutral colors while  \CQ $B$ is not
significantly more red than \CQ $A$.  Future observations that reduce
the photometric uncertainties will be required to reach stronger
conclusions.

\subsection{Frequency of Binaries}

We have observed 75 TNOs so far in our photometric survey
using HST and have examined all of these for possible binary
companions.  Of those 75, we have detected companions to 3, \CQ , \CF ,
and the previously known binary \WW .  This statistic applies to
companions with separations greater than $\sim$0.15 arcsec and V-band
differences of less than $\sim$1 mag.   Although \WW\ was identified as
a binary prior to its inclusion in our target list, the magnitude of
\WW\ and the precision of its orbit determination matched the criteria
for inclusion and it would have been added even if
the binary were not known.  Also, the order in which snapshot targets
are scheduled is random and outside the control of the observers.  The
probablility that \WW\ would have been observed at this point in our
observing program is approximately 50\%.  Given these considerations,
we believe it is proper to consider \WW\ as part of an unbiased sample.
 We then conclude that the frequency of binaries consistent with the
limits above is simply 3/75 or $\approx 4\pm 2$\%.   We note that,
because of the large statistical uncertainty, the exclusion of \WW\
from the sample would not alter this conclusion substantially.

A frequency of $\approx 4\pm 2$\% is in marginal agreement with the
detection of 8 TNBs out of a total of more than 500 TNOs detected to
date.  However, the uneven sampling and data quality of this larger
population complicates any statistical conclusions from this larger
sample.  Many of the objects have been observed only a few times and
some are effectively lost.  Many of the detected binaries, \CQ\ for
example, could not be detected by most ground-based systems including
those equipped with adaptive optics. \CF\ is just detectable with
ground-based systems, and indeed has been verified as binary from the
ground (Romanishin \& Tegler et al. 2002).  Furthermore, the magnitude
limit and angular resolution in individual observations of TNOs is a
strong function of equipment and observing conditions.  Nevertheless,
the overall detection rate of $\approx 1.5\pm 0.5$\% supports the rate
inferred from our more uniform survey.  

It is worth noting that half of known TNBs, four of eight, have
been discovered with HST (Trujillo and Brown 2002, Brown and Trujillo
2002, Noll and Stephens 2002, Noll et al. 2002).  Since far more
observations of TNOs are made by groundbased telescopes than by HST,
this fact would appear to imply that TNBs separated by more than an
arcsecond are less frequent than more closely spaced companions, an
assertion borne out by Table 1 when one considers that 1 arcsec at 40 AU
is a physical distance of 29,000 km.

We also note that all three of the binaries detected by us and five of
the eight objects in Table 1 are classical TNOs.  Two are in the 3:2
resonance and one is possibly a scattered object.  Of the 75 objects
observed in our program so far, 65\% are classical TNOs with low
eccentricity and inclination, 24\% are plutinos, and 11\% are
scattered disk objects.  There does not seem to be sufficient evidence
at present to make any conclusions about the relative frequency of
binaries in the different dynamical populations, though eventually this
information may be useful in determining the origin and history of TNBs
and the dynamical classes in the Kuiper Belt.

Finally, it is worthwhile to make a comparison between TNBs and
binaries found in the main belt and near-Earth populations.  A total of
6 binaries have been identified out of a sample of 300 main-belt
targets by Merline et al.~(2001), a rate, at first glance, roughly
similar to that being found in the Kuiper Belt.  However, it must be
noted that the relative sizes and separations of main belt and Kuiper
belt binaries appear to be significantly different (Margot 2002).  If a
population of TNBs with faint and close-in companions, similar to the
main belt binaries, exists, the fraction of TNBs could be significantly
higher than our current estimate. If collisions are the main source of
binaries in both populations, the frequency of binaries, and their
orbital and size distribution can be used to probe the efficiency of
binary formation from such events.   The near-Earth asteroid (NEA)
population has a higher frequency of binaries, as high as 16\% for
objects over 200 m in diameter (Margot et al.~2002).  However,
mechanisms other than collisions, e.g. tidal breakup, may be
responsible for this larger number among the NEAs and this population
may be less relevant for comparison to TNBs.

\bigskip\bigskip

\noindent{\bf References} 

\ref {Baggett, S. et al. 2002, in HST WFPC2 Data Handbook, ed.
B.~Mobasher, Baltimore, STScI}

\ref {Biretta, J., et al. 2001, WFPC2 Instrument Handbook, Baltimore, STScI}

\ref {Brown, M.~E. \& Trujillo, C.~A. 2002, IAUC 7807}

\ref {Buie, M.~W., Cruikshank, D.~P., Lebofsky, L.~A., \& Tedesco,
E.~F.  1987  Nature 329, 522}

\ref {Buie, M.~W., Tholen, D.~J., \& Horne, K. 1992, Icarus 97, 211}

\ref {Chen, J., Trujillo, C., Luu, J., Jewitt, D., Kavelaars, J.~J.,
Gladman, B., \& Brown, W.  1997 MPEC 1997-J02}

\ref {Christy, J.~W. \& Harrington, R.~S. 1978, AJ 83, 1005}

\ref {Dolphin, A. 2000, PASP 112, 1397}

\ref {Eliot, J.~L. 2001, IAUC 7733}

\ref {Fink, U. \& DiSanti, M.  1988, AJ 95, 229}

\ref{Jewitt, D. \& Luu, J. 2001, AJ 122, 2099}

\ref {Kavelaars, J.~J., Petit, J.-M., Gladman, B., \& Holman, M. 2001,
IAUC 7749}

\ref {Krist, J.~E. \& Hook, R.  1997, {\it TinyTim User's Manual}, v.4.4,
Space Telescope Science Institute: Baltimore}

\ref{Luu, J., \& Jewitt, D. 1996, AJ 112, 2310}


\ref {Margot, J.~L. 2002, Nature 416, 694}

\ref {Margot, J.~L., Nolan, M.~C., Benner, L.~A.~M., Ostro, S.~J.,
Giorgini, J.~D., Slade, M.~A., \& Campbell, D.~B. Science Online, April
11, 2002}

\ref {Merline, W.~J., Close, L.~M., Menard, F., Dumas, C., Chapman,
C.~R., Slater, D.~C. 2001, BAAS 33, \#52.01}

\ref {Millis, R.~L., Buie, M.~W., Wasserman, L.~H., Elliot, J.~L.,
Kern, S.~D., and Wagner, R.M. 2002, AJ 123, 2083}

\ref {Noll, K. \& Stephens, D. 2002, IAUC 7824}

\ref {Noll, K., Stephens, D., Grundy, W., Spencer, J., Millis, R., Buie,
M., Cruikshank, D., Tegler, S., \& Romanishin, W. 2002b, IAUC 7857}

\ref {Romanishin, W. \& Tegler, S.~C. 1999, Nature 398, 129}

\ref {Romanishin, W., Tegler, S., Noll, K. Stephens, D, Grundy, W.,
Spencer, J., Millis, R., Buie, M., \& Cruikshank, D. 2002, IAUC 7962}

\ref {Trujillo, C.~A., \& Brown, M.~E. 2002, IAUC 7787}

\ref {Whitmore, B., Heyer, I., \& Casertano, S. 1999, PASP 111, 1559}

\ref {Veillet, C. 2001, IAUC 7610}

\ref {Veillet, C., Parker, J.~W., Griffin, I., Marsden, B.,
Doressoundiram, A., Buie, M., Tholen, D.~J., Connelley, M., \& Holman,
M.~J. 2002 Nature }

\ref {Young, E.~F., Galdamez, K., Buie, M.~W., Binzel, R.~P., \& Tholen,
D.~J. 1999, AJ 117, 1063}

\newpage

{\tenrm
\null\vskip .1in
\tabskip=1.5em 
\baselineskip=12pt
\tolerance=10000
$$\vbox{ \halign {
#\hfil & #\hfil & #\hfil & #\hfil & #\hfil  \cr
\multispan5\hfil{\bf Table 1: Trans-Neptunian Binaries}\hfil \cr
\noalign { \vskip 12pt \hrule height 1pt \vskip 1pt \hrule height 1pt \vskip 8pt } 
object & dynamical  & separation  & $\Delta$ m  & reference  \cr
       & class      &   (km)$^*$  &             &            \cr
\noalign { \vskip 8pt\hrule height 1pt \vskip 8pt }
\noalign {\bigskip } 
Pluto/Charon  & plutino    & 19,366 &  1.4   &  Christy \& Harrington (1978)  \cr
\noalign {\smallskip}
\WW           &  classical & 22,300 &  0.4  &  Veillet (2001)  \cr
\noalign {\smallskip}
\QT           & classical  & 19,000 &  0.55 &  Elliot (2001)  \cr
\noalign {\smallskip}
\QW           & classical  &126,000 &  0    &  Kavelaars et al. (2001)  \cr
\noalign {\smallskip}
\TC           & plutino    &  8,300 & 1.9   &  Trujillo \& Brown (2002)  \cr
\noalign {\smallskip}
\SM           & scattered  &  5,600 &  1.9  &  Brown \& Trujillo (2002)  \cr
\noalign {\smallskip}
\CQ           & classical  &  8,100 &  0.24 &  Noll \& Stephens (2002)  \cr
\noalign {\smallskip}
\CF           & classical  & 23,360 & 0.87  &  Noll et al. (2002)  \cr
\noalign {\smallskip}
\noalign {\vskip 8pt \hrule height 1pt }  
\multispan5{$^*$ Semi-major axis listed for Pluto/Charon and \WW .}\hfil\cr
\multispan5{\ \   For other objects, mean separation at time of observations is listed. }\hfil \cr 
  } }$$}

\newpage

{\tenrm
\null\vskip .1in
\tabskip=1.5em 
\baselineskip=12pt
\tolerance=10000
$$\vbox{ \halign {
#\hfil & #\hfil  & #\hfil & #\hfil \cr
\multispan4\hfil{\bf Table 2: Positional Data}\hfil \cr
\noalign { \vskip 12pt \hrule height 1pt \vskip 1pt \hrule height 1pt \vskip 8pt } 
date & object &  separation  & position angle \cr
\noalign { \vskip 8pt\hrule height 1pt \vskip 8pt }
\noalign {\bigskip } 
 2001 Nov 17.29 & \CQ\ & 0.20$\pm$0.03    & 16$\pm$4 \cr
\noalign {\smallskip}
 2002 Jun 18.33 &      & 0.337$\pm$0.010  & 334.0$\pm$1.2 \cr
\noalign {\smallskip}
 2002 Jun 30.15 &      & 0.334$\pm$0.004  & 337.9$\pm$0.9 \cr
\noalign {\smallskip}
 2002 Jul 12.17 &      & 0.331$\pm$0.010  & 340.7$\pm$0.8 \cr
\noalign {\bigskip}
 2002 Jan 12.06 & \CF\ & 0.78$\pm$0.03    & 106.6$\pm$2.5 \cr
 2002 Apr 11.3 &      & 0.8$\pm$0.2      & 103$\pm$5 \cr
\noalign {\vskip 8pt \hrule height 1pt }  
  } }$$}   

\newpage

{\tenrm
\null\vskip .1in
\tabskip=1.5em 
\baselineskip=12pt
\tolerance=10000
$$\vbox{ \halign {
#\hfil & #\hfil & #\hfil & #\hfil & #\hfil & #\hfil & #\hfil \cr
\multispan7\hfil{\bf Table 3: Photometry}\hfil \cr
\noalign { \vskip 12pt \hrule height 1pt \vskip 1pt \hrule height 1pt \vskip 8pt } 
date & object & component &  V  & R & I & V-I  \cr
\noalign { \vskip 8pt\hrule height 1pt \vskip 8pt }
\noalign {\bigskip } 
 17 Nov 2001 & \CQ\ &A &  23.78$\pm$0.13 &23.27$\pm$0.19 &22.54$\pm$0.09 &1.24$\pm$0.16 \cr
             &      &B & 24.02$\pm$0.22 &23.54$\pm$0.23 &23.06$\pm$0.18 &0.9$\pm$0.3 \cr
\noalign {\smallskip}
 18 Jun 2002 &  &A &           &               &23.42$\pm$0.04  & \cr
             &      &B &           &               &23.03$\pm$0.03  & \cr
\noalign {\smallskip}
 30 Jun 2002 &  &A &           &               &22.63$\pm$0.03  & \cr
             &      &B &           &               &22.96$\pm$0.04  & \cr
\noalign {\smallskip}
 12 Jul 2002 &  &A &           &               &22.69$\pm$0.03  & \cr
             &      &B &           &               &23.06$\pm$0.04  & \cr
\noalign {\bigskip}
12 Jan 2002  & \CF\ &A &24.25$\pm$0.11 &  & 22.99$\pm$0.08 & 1.26$\pm$0.14 \cr
             &      &B &25.12$\pm$0.24 &  & 23.55$\pm$0.13 & 1.6$\pm$0.3 \cr
\noalign {\vskip 8pt \hrule height 1pt }  
  } }$$}

\begin{figure}
\epsfxsize=6.0in
\epsffile{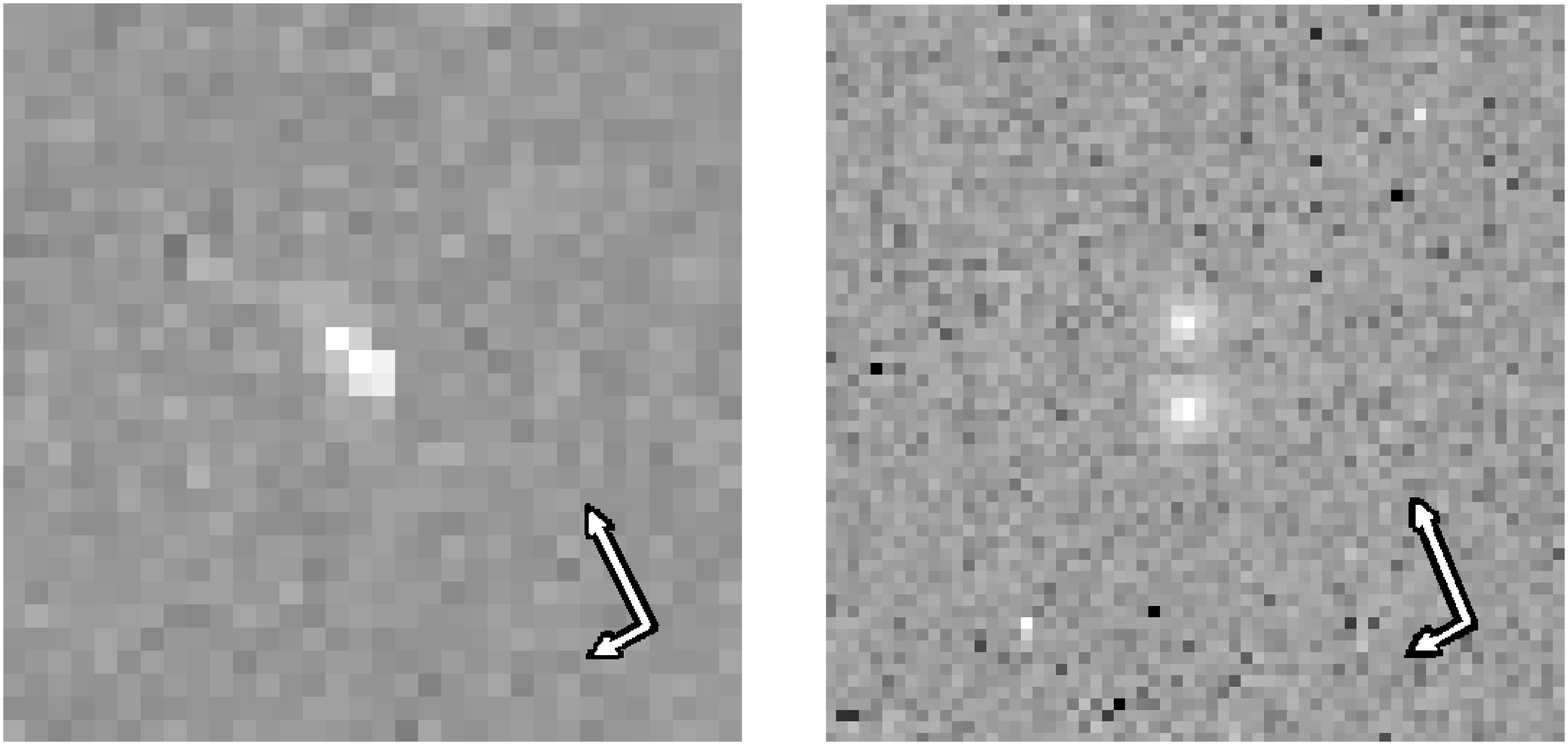}

\figurecaption {{\bf Figure 1.} a) The image shown on the left
is a combination of two individual 160 second integrations in the F814W
filter taken on 17 November 2001.  We show a portion of the WF3 chip
around the location of \CQ .  The elongation of \CQ\ is apparent, though
the individual components are not clearly resolved.  b) The image on the
right is a combination of three 800 second images taken on 30 June 2002
through the F814W filter. The image shown is a 30 arcsecond portion of
the PC chip centered on \CQ .  In this image the binary is clearly
resolved.  The long arrow indicates North on each panel and the short
arrow, East.}

\end{figure}

\begin{figure}
\epsfxsize=6.0in
\epsffile{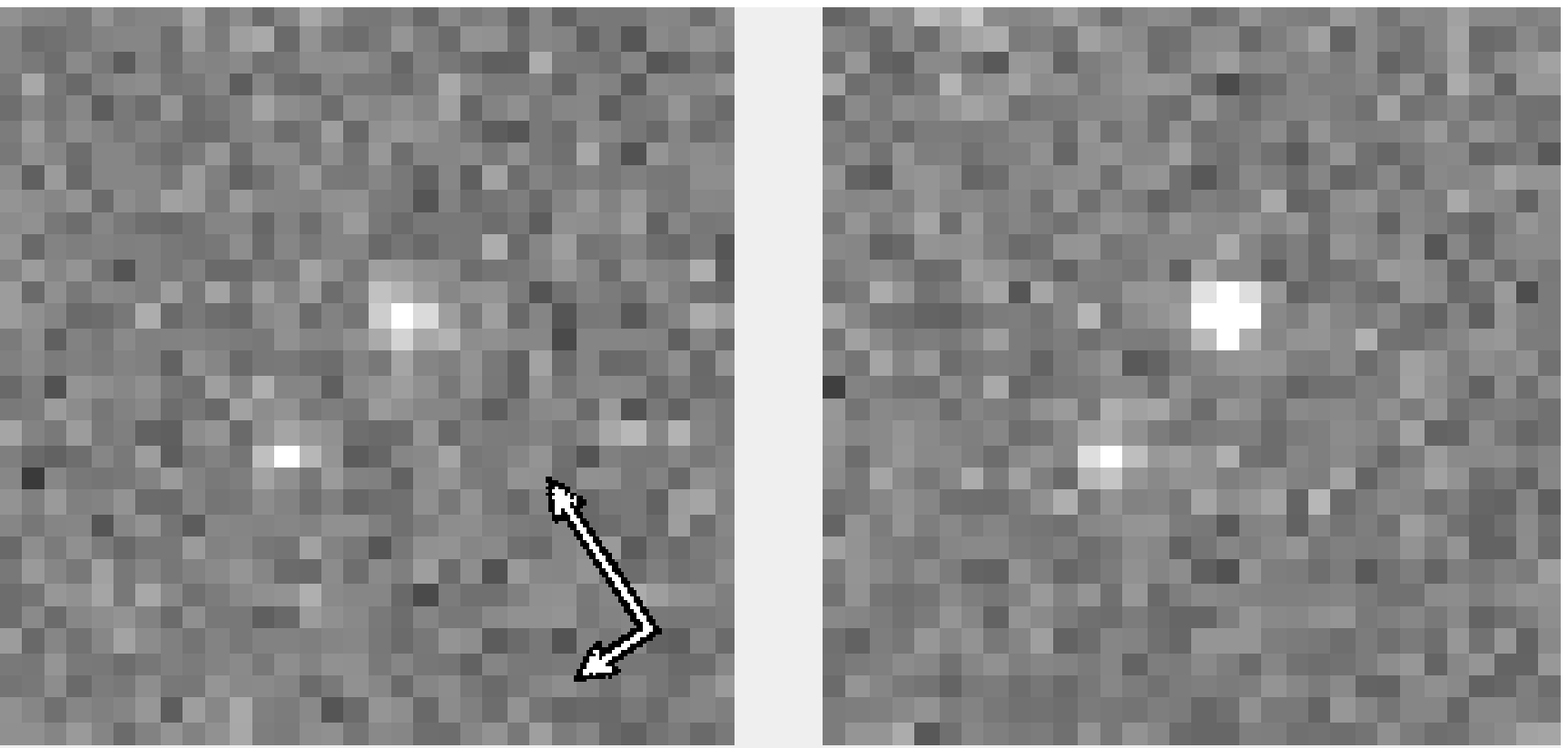}

\figurecaption {{\bf Figure 2.} Left to right, F555W, and F814W
images of \CF\ taken with WFPC2 on 12 January 2002.  The images shown
are combinations of two individual 260 second integrations in each
filter as described in the text.  We show a portion of the WF3 chip
around the location of \CF .  The two components are cleanly separated
and are visible in both filters. North and east are indicated by the
long and short arrows respectively. }

\end{figure}

\begin{figure}
\epsfxsize=5.0in
\epsffile{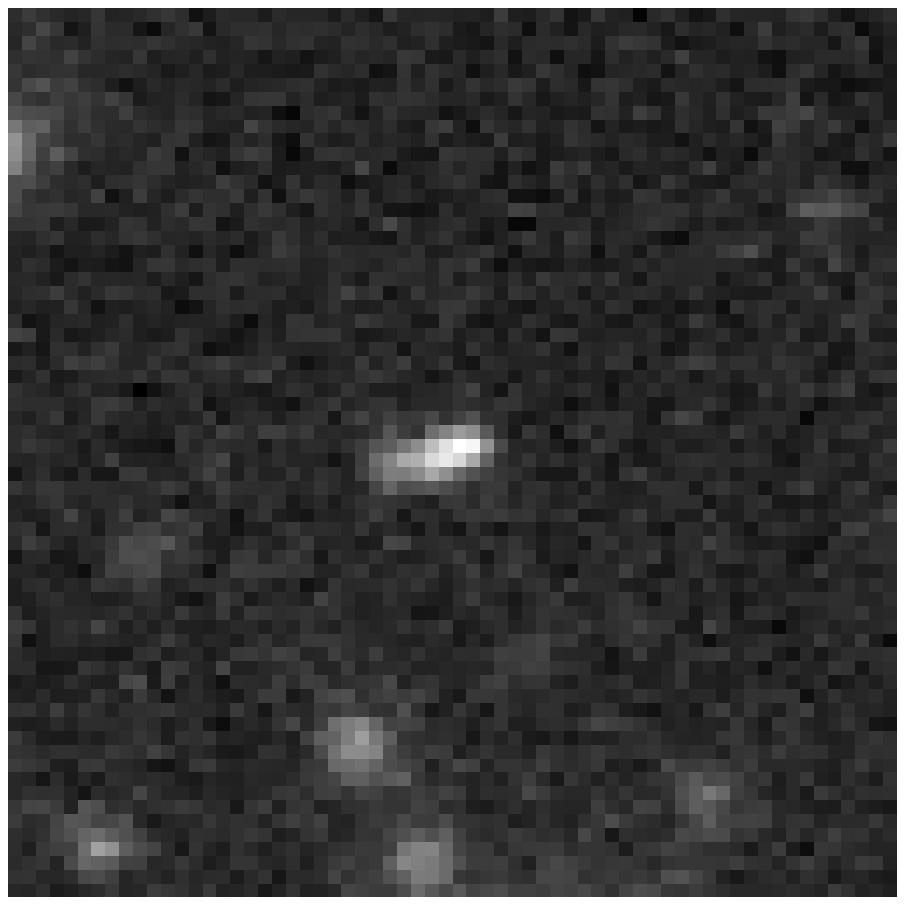}

\figurecaption {{\bf Figure 3.}  Image of \CF\ obtained with LRIS at
the Keck I telescope on 11 April 2002.  The individual images have been
shifted and registered with offsets computed to match the small motion
of \CF .  North is up and east to the left in this image.  Motion
during the individual integrations was less than 0.2 pixels.  Pixels
are 0.215 arcsec on a side.  The object is clearly elongated in a
direction that matches the two resolved components in the HST image,
confirming the existence of the binary companion.}

\end{figure}
 
\begin{figure}
\epsfxsize=6.0in
\epsffile{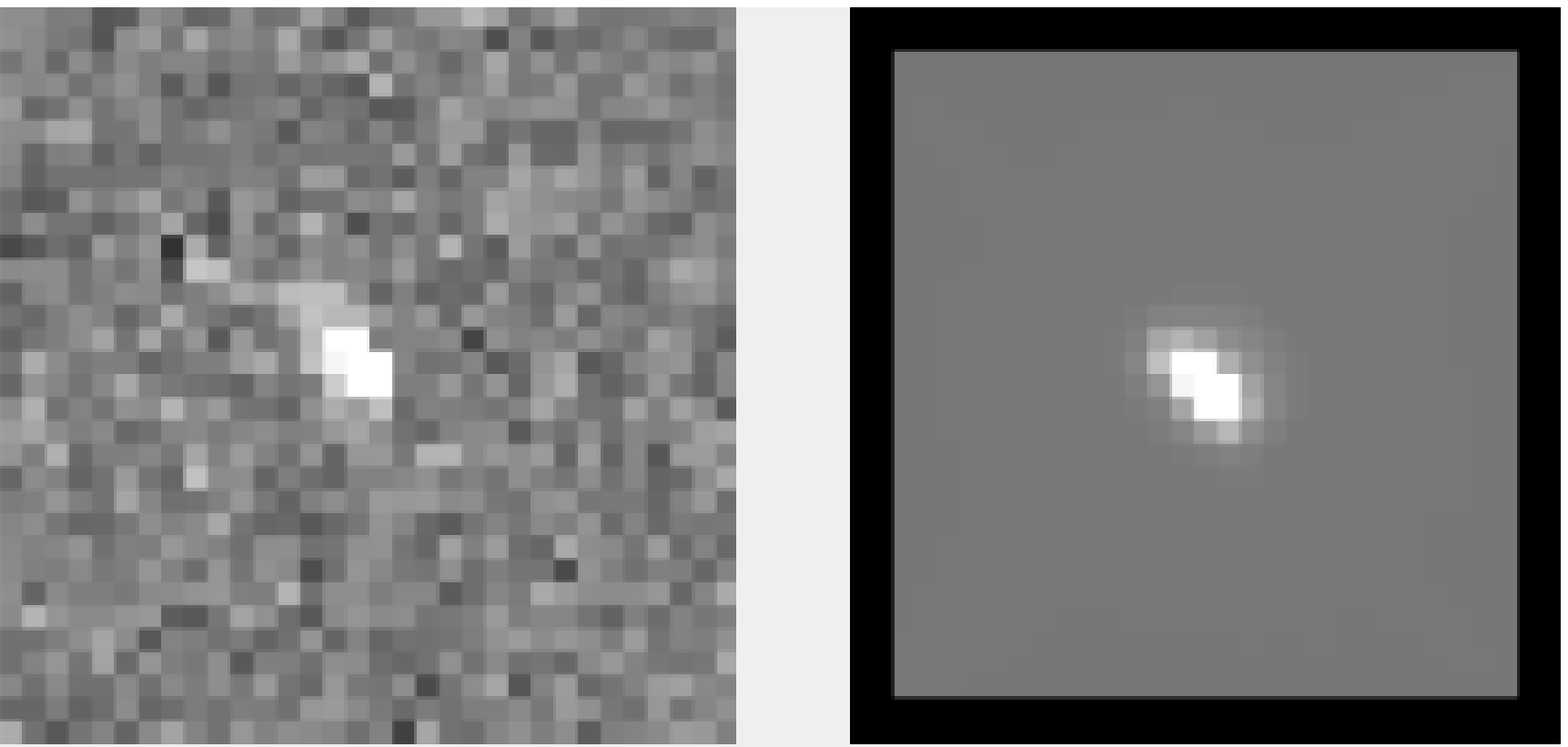}

\figurecaption {{\bf Figure 4.} a) The F814W image of \CQ\ and b) a
best-fit model created with synthetic PSFs combined as described in the
text.   }

\end{figure}

\end{document}